\documentclass[sigconf]{aamas}

\usepackage{balance} 
\usepackage{amsmath}
\usepackage{algorithm}  
\usepackage{algpseudocode}
\usepackage{subfigure}
\usepackage{subcaption}  
\usepackage{booktabs}  
\usepackage{multirow}  
\graphicspath{{images}}  

\setcopyright{ifaamas}
\acmConference[AAMAS '25]{Proc.\@ of the 24th International Conference
on Autonomous Agents and Multiagent Systems (AAMAS 2025)}{May 19 -- 23, 2025}
{Detroit, Michigan, USA}{A.~El~Fallah~Seghrouchni, Y.~Vorobeychik, S.~Das, A.~Nowe (eds.)}
\copyrightyear{2025}
\acmYear{2025}
\acmDOI{}
\acmPrice{}
\acmISBN{}

\acmSubmissionID{98}

\title[AAMAS-2025 Formatting Instructions]{MacLight: Multi-scene Aggregation Convolutional Learning for Traffic Signal Control}

\author{Sunbowen Lee}
\affiliation{
  \institution{College of Science\\ Wuhan University of Science and Technology}
  \city{Wuhan}
  \country{China}}
\email{bw1863@wust.edu.cn}

\author{Hongqin Lyu}
\authornote{Contribution equal to the first author}
\affiliation{
  \institution{SKLP, ICT, CAS\\ University of Chinese Academy of Sciences}
  \city{Beijing}
  \country{China}}
\email{hongqinlyu@163.com}

\author{Yicheng Gong}
\authornote{Corresponding author}
\affiliation{
  \institution{College of Science\\ Wuhan University of Science and Technology}
  \city{Wuhan}
  \country{China}}
\email{gongyicheng@wust.edu.cn}

\author{Yingying Sun}
\affiliation{
  \institution{College of Science \\ Wuhan University of Science and Technology}
  \city{Wuhan}
  \country{China}}
\email{sunyingying082@163.com}

\author{Chao Deng}
\affiliation{
  \institution{School of Automobile and Traffic Engineering, Wuhan University of Science and Technology}
  \city{Wuhan}
  \country{China}}
\email{woec@wust.edu.cn}

\begin{abstract}
  Reinforcement learning methods have proposed promising traffic signal control policy that can be trained on large road networks. Current SOTA methods model road networks as topological graph structures, incorporate graph attention into deep Q-learning, and merge local and global embeddings to improve policy. However, graph-based methods are difficult to parallelize, resulting in huge time overhead. Moreover, none of the current peer studies have deployed dynamic traffic systems for experiments, which is far from the actual situation.

  In this context, we propose Multi-Scene Aggregation Convolutional Learning for traffic signal control (MacLight), which offers faster training speeds and more stable performance. Our approach consists of two main components. The first is the global representation, where we utilize variational autoencoders to compactly compress and extract the global representation. The second component employs the proximal policy optimization algorithm as the backbone, allowing value evaluation to consider both local features and global embedding representations. This backbone model significantly reduces time overhead and ensures stability in policy updates. We validated our method across multiple traffic scenarios under both static and dynamic traffic systems. Experimental results demonstrate that, compared to general and domian SOTA methods, our approach achieves superior stability, optimized convergence levels and the highest time efficiency. The code is under \url{https://github.com/Aegis1863/MacLight}.
\end{abstract}


\newcommand{\BibTeX}{\rm B\kern-.05em{\sc i\kern-.025em b}\kern-.08em\TeX}

\begin{document}


\pagestyle{fancy}
\fancyhead{}


\maketitle


\section{Introduction}

Traffic signal control (TSC) is an important issue in urban management. As the number of vehicles owned by residents increases, the deteriorating traffic conditions have a serious impact on social development. Traffic signal optimization is a low-cost means to alleviate traffic pressure.

The optimization of traffic light timing constitutes a complex nonlinear stochastic problem, as highlighted in \cite{5978226}. Traditional intelligent control solutions often resort to assumptions or lack of flexibility, such as unlimited lane capacity \cite{Varaiya2013}, Christina Diakaki et al. \cite{DIAKAKI2002183} assumes that the traffic flow is uniform, or fail to adapt effectively to dynamic traffic flows \cite{gershenson2005selforganizingtrafficlights}. Consequently, the performance may fall short of that achieved by a fixed timing plan meticulously crafted by human experts.

Although mathematical modeling of real traffic systems is very difficult, the emergence of mature traffic simulators can provide interactive environments, which means that model-free methods can be applied. Reinforcement learning (RL) \cite{Sutton1988, sutton1999, sutton2018reinforcement} provides SOTA solutions in the field of model-free control. Preliminary RL approaches, such as Q-learning \cite{Watkins1992} and its variants, have shown promising results in optimizing TSC. By iteratively learning from the environment, these algorithms can dynamically adjust signal to minimize congestion.

Intelligent control methods for individual traffic lights are very mature \cite{presslight}, but they are inefficient for large road networks. Current research focuses on whether multiple traffic lights can effectively coordinate to achieve effects such as green wave roads. A common approach is to model the road network as a topology graph structure and introduce Graph Attention Networks (GAT) \cite{velickovic2018graph}, enabling traffic signals to consider both local and neighboring features for comprehensive optimization decisions through feature aggregation. However, current GAT-based approaches \cite{colight,DuaLight, Metalight} are almost used in Deep Q-learning (DQN) \cite{DQN}. DQN, As an off-policy framework, despite being data-efficient, graph learning and batch learning consume significant time and computational resources. A more critical issue is they are prone to overfitting, leading to policy collapse.

In this case, we consider both local and global characteristics and propose a novel global scene aggregation approach.
\textbf{Our approach is motivated by two key points:} \textbf{firstly, the ability of decision-making and value evaluation of agents should to be separated.} Thus, we utilize Proximal Policy Optimization (PPO) \cite{PPO} as the backbone model, which has a value evaluation module and a policy improvement module to process different information respectively.
\textbf{Secondly}, global scene aggregation does not necessarily require topological graph modeling. Research by Hua Wei et al. \cite{colight} indicates that in topological graph modeling scenarios, \textbf{each agent considering only one-hop neighbors yields the best results, which limits the agent's understanding of broader states.} Therefore, we aggregate the features of each agent (scene), using convolutional neural networks (CNN) \cite{NIPS2012c399862d} for a latent global representation. Another important reason for not using graph convolutional neural networks (GCNs) is that it is difficult to compute in parallel and apply to the more advanced Actor Critic RL framework. Consequently, our approach is called multi-scene aggregation convolutional learning (MacLight).

Furthermore, we are the first construct \textbf{dynamic traffic flow scenario} by using the professional open-source simulator SUMO \cite{SUMO2018}. It can simulate the change of traffic flow distribution caused by emergency traffic incidents. We incorporate it as a challenging experimental scenario, alongside other general scenarios to test algorithms. Specifically, we can impose emergency speed limits or ban traffic on any road and reroute all vehicles. Vehicles will consider speed limits or prohibitions and choose new routes, leading to sudden changes in traffic distribution on other roads. It requires agents not only to cope with familiar traffic characteristics but also to have the ability to handle dynamically changing traffic flows. This greatly expands the scope of existing research.

In summary, the contributions of this paper are as follows:
\begin{itemize}
  \item[1.] We construct a dynamic traffic flow simulation scheme to simulate any possible emergency traffic event, greatly expanding the current research space.
  \item[2.] We propose an online-trained variational autoencoder (VAE) based on CNN for global state representation, obtaining a compact and efficient representation from the latent space for downstream learning.
  \item[3.] We integrate global state representation into the value evaluation module of PPO, enabling the algorithm to balance local and macro characteristics, and demonstrating superior performance compared to both general and domain SOTA.
\end{itemize}


\section{Related work}

\textbf{Customizable simulator.} In the field of TSC, the Simulation of Urban MObility (SUMO) simulator is widely used for urban planning and traffic flow simulation. The simulator allows researchers to define any desired traffic flow scenario. Ma and Wu \cite{SUMOearly} were among the first to utilize SUMO for traffic control simulations, and it has become the main tool for relevant researchers in recent years. Furthermore, SUMO-RL \cite{sumorl, Alegre2021} integrates SUMO with the OpenAI Gym environment, facilitating RL training in TSC.

\textbf{Intelligent traffic control.} In a multi-agent system, domain knowledge becomes a key for communication and coordination between agents. Some early methods such as PressLight \cite{presslight} have achieved good single-agent control and proposed feasible training methods. MPLight \cite{Thousand} is based on PressLight and extends it to large road networks, using the same model to make decisions for all intersections, which requires that the state space and action space of each intersection are consistent. After Afshin Oroojlooy et al. \cite{AttendLight} introduced the attention mechanism into this field, the GAT method gradually became mainstream. From a multi-agent perspective, several traffic lights are usually regarded as multiple different agents, and the road network is regarded as a topological structure to model the data structure. In this case, GAT becomes the main optimization method. For example, CoLight \cite{colight} is based on the DQN method and uses GAT to assign weights to neighbors. Experiments show that each agent works best when it only pays attention to itself and its one-hop neighbors. STMARL \cite{9240060} and DynSTGAT \cite{DynSTGAT} use a LSTM \cite{LSTM} or TCN \cite{TCN} to capture historical state information (such as traffic flow) and use a graph convolutional network (GCN) or GAT to obtain spatial dependencies. DuaLight \cite{DuaLight} introduces scene characteristics based on CoLight, introduces neighbor weighted matrices and feature-weighted matrices for each agent, and also performs GAT representation on the one-hop neighbors, further enhancing the agent's understanding of its own scene and local coordination capabilities. GuideLight \cite{GuideLight} implements a control method that is closer to industrial needs based on cyclic phase switching and combined with behavioral cloning and curriculum learning training models.


\begin{figure}[t]
  \centering
  \includegraphics[width=\linewidth]{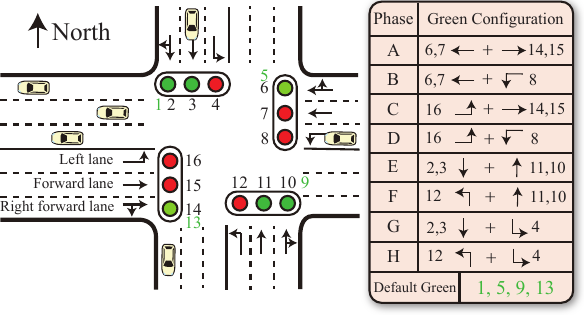}
  \caption{General right-hand 2-way 6-lane intersection with eight non-conflicting green light signal configurations. The right turn lane signal is green by default. Traffic signal numbers start from the north and go clockwise.}
  \label{fig:phase_demo}
  \Description{An intersection with eight non-conflicting green light signal configurations.}
\end{figure}

\section{Notation}

We define the key concepts in RL for TSC before introducing our model, including the signal configuration and modeling TSC as a Partially Observable Markov Decision Process (POMDP).

\textbf{Intersection.} Fig. \ref{fig:phase_demo} shows a general right-hand 2-way 6-lane intersection. We define the traffic light numbers starting from the north and proceeding clockwise. The left-turn lanes can only be used for left turns, while the right-turn lanes can be used for both going straight and turning right. Among these, traffic lights numbered 1, 5, 9, and 13 are for right-turn lanes and are default to green. However, when going straight in a right-turn lane, one must obey the green signal of going straight. Each time 3 or 4 green light signals are given to allow passing, a total of 8 signal combinations are defined in the scenario.

\textbf{Phase.} Referring to the table on the right side of Fig. \ref{fig:phase_demo}, we define an equitable signal configuration scheme that ensures no lane conflicts exist for any passing scenarios within a single cycle and each lane has two opportunities for passing within the cycle. This scheme is consistent with most real-world configurations, and the algorithm can adapt individually even if there are different configuration schemes.

\textbf{POMDP Modeling for TSC.} The traffic signal control problem is modeled as a POMDP. We consider each intersection as an independent agent that faces continuously changing traffic conditions and can only observe its own information completely, without grasping the global state. Another principle is that the next state is only affected by the current state and the current decision, and has nothing to do with the previous state. A POMDP can be described by a tuple $\langle \mathcal{S}, \mathcal{O}, \mathcal{A}, \mathcal{P}, \mathcal{R}, \mathcal{\pi}, \mathcal{\gamma}, \rangle$ and are introduced below.

\textbf{Global state space $\mathcal{S}$ \& Partial state space $\mathcal{O}$}. The partial observation of agent $i$ at time $t$ is $o^t_i\in \mathcal{O}$, while global state $s^t \in \mathcal{S}$ and $o^t_i \in s^t$. Partial observations are also called local observations in following context. Refer to \cite{Alegre2021}, each local observation consists of four parts:
\begin{enumerate}
    \item[1.] The current action represented as a one-hot vector;
    \item[2.] A boolean value indicating whether the current signal allows switching. We specify that each action must remain in place for at least 10 seconds to meet real-world requirements;
    \item[3.] The vehicle density in each lane, calculated as the number of vehicles in the lane divided by the lane capacity;
    \item[4.] The density of waiting vehicles in each lane, calculated as the number of stopped vehicles divided by the lane capacity;
\end{enumerate}
These components are encoded into a vector to represent the current state of each intersection.

\textbf{Action $\mathcal{A}$.} In the case of Fig. \ref{fig:phase_demo}, the eight phases correspond to eight different action choices. At time $t$, the action of agent $i$ is $a^t_i \in \mathcal{A}$. In the simulation, by default, we provide the corresponding yellow signal before switching the red signal.

\textbf{Transition probability $\mathcal{P}$.} Due to the Markov property, the probability transfer function is expressed as $\mathcal{P}(s^{t+1}|s^t,a^t)$. The specific form of the function is unknown and is usually represented by reality or a simulator. We perform RL to capture the dynamic characteristics.

\textbf{Reward $\mathcal{R}$.} Referring to the design of Alegre et al. \cite{Alegre2021}, we first define the waiting time of the vehicle. At time $t$, the total waiting time of all vehicles stopped at intersection (agent) $i$ is denoted as $W^t_i$. Then the agent's reward is $r^t_i=W^{t-1}_i-W^{t}_i$. Our goal is to maximize the reward, which means that the agent should try to make the current waiting time shorter than the previous waiting time. The final reward is expected to converge to around 0, that is, the system reaches a state of equilibrium. The advantage of considering waiting time as a reward is that the agent will not deliberately delay the release time of some lanes due to fewer cars there, but instead balanced take all vehicles into consideration.

There are many reward functions. For example, the reward value can increase with the decrease in the number of blocked vehicles or set a pressure indicator \cite{presslight} to measure the difference between the number of vehicles entering and leaving the lane. We test various reward functions in "ingolstadt21" \cite{ault2021reinforcement}, and this scenario is completely different from ours. We adopt the same independent PPO (IPPO) for all experiments. In this case, we evaluate various indicators and determine that the aforementioned method is the best choice, with superior performance compared to other methods. The experimental results are shown in Table \ref{tab:comparison}.

\begin{table}[t]
  \centering
  \caption{Comparison of different reward methods. The first column includes various reward targets and a baseline, and the first row is the system indicators. Arrows indicate the better direction, and standard deviations in brackets are obtained from multiple experiments. The fixed time is similar to the real-life solution, that is, the fixed time switching signal, which is a baseline.}
  {\small\begin{tabular}{lccc}
  \toprule
  \textbf{}        & \textbf{Waiting↓} & \textbf{Queue↓} & \textbf{Speed↑} \\
  \midrule
  Pressure         & 2106.7 (1283)     & 40.8 (7)        & 7.9 (0.4)       \\
  Queue            & 4358.5 (2608)     & 50.4 (8)        & 7.5 (0.3)       \\
  Speed            & 1009.9 (597)      & 31.5 (9)        & 8.4 (0.4)       \\
  Waiting          & 790.1 (703)       & 23.8 (10)       & 8.8 (0.6)       \\
  Fixed time      & 684.8             & 70.2            & 8.1             \\
  \midrule
  Our adoption  & \textbf{422.0 (577)} & \textbf{21.9 (11)} & \textbf{9.0 (0.6)} \\
  \bottomrule
  \end{tabular}}
  \label{tab:comparison}
\end{table}

\textbf{Policy $\mathcal{\pi}$.} The decision made by agent $i$ in time $t$ based on the current partial observation $o^t_i$ is given by the policy function $\pi^t_i(a^t_i|o^t_i)$. The agent policy should maximize the total reward $\sum^T_{t=\tau}\gamma^{t-\tau}r^t_i$, where $\gamma$ is the discount factor, usually 0.98. This means that agents discount future reward and care first about near-term reward.

\begin{figure*}[h]
  \centering
  \includegraphics[width=\linewidth]{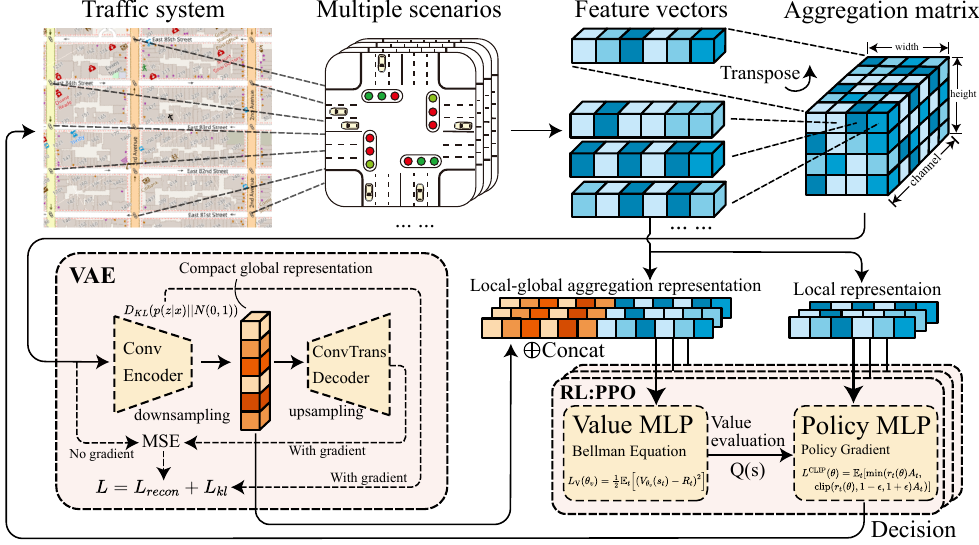}
  \caption{MacLight framework. The first row shows how to construct the aggregation matrix, and the second row introduces the main model frameworks, including VAE, PPO.}
  \label{fig:total_framework}
  \Description{Multi-scenario aggregation convolutional learning for traffic signal control.}
\end{figure*}

\section{Methodology}
In this section, we will introduce the implementation of MacLight, including information aggregation, VAE feature compression, PPO and method of dynamic traffic flow construction.

\subsection{Multi-scene aggregation matrix}

Considering the geographical invariance of intersections, we organize the global information into a three-dimensional matrix. The global information is obtained by merging several local information, and each local information can be regarded as a scene. The local information of each intersection can be represented as a feature vector. Each vector is appropriately transposed and organized according to its location in geographic space, ultimately forming a high-dimensional global feature matrix as shown in the upper right of Fig. \ref{fig:total_framework}. The width and height of the matrix correspond to the geographical locations, and the number of channels is equal to the length of a single feature vector.

Clearly, the grid-based setting is the basis for adopting CNN as a representation model. Although real-world road networks do not appear as regular as pixel grids, considering that most intersections are four-armed, the grid-like characteristics can still be observed when transforming them into a graph.

\subsection{Autoencoder}

For feature extraction of three-dimensional matrices, we construct a VAE based on CNN. The structure refers to the VAE in Fig. \ref{fig:total_framework}. The encoder performs downsampling and finally outputs a compact compressed representation. The decoder restores the representation to the original matrix. Its training is carried out according to the method of Diederik P. Kingma and Max Welling \cite{VAE}.

We first give the process of upsampling, for a matrix $x$ with a channel length of 33, the process is
\begin{equation}
  h=\mathrm{Conv}_3^{256}\big[\mathrm{ReLU}(\mathrm{Conv}_3^{128}(\mathrm{ReLU}(\mathrm{Conv}_3^{64}(x))))\big].
\end{equation}

The parameters of the Gaussian distribution represented in the latent space are calculated as
\begin{equation}
\begin{aligned}
  \mu&=W_\mu h+b_\mu,\\
  \log\sigma^2&=W_{\text{logvar}}h+b_{\text{logvar}},
\end{aligned}
\end{equation}
where $W_{\mu}$, $b_\mu$, $W_{\text{logvar}}$ and $b_{\text{logvar}}$ correspond to weights and biases respectively. Then, an effective compact representation $z$ is obtained by Gaussian distributions built on $\mu$ and $\sigma$:
\begin{equation}
  z=\mu+\epsilon\cdot\sigma,\quad\epsilon\sim\mathcal{N}(0,I).
\end{equation}

The decoder uses transposed convolution models:
\begin{equation}
  \begin{aligned}
  z_{reshape}=&~\text{Reshape}(\text{Linear}(z)),\\
  x_\text{recon}=&~\text{Sigmoid}\big[\text{ConvTrans}_3^{33}\big[\text{ReLU}(\text{ConvTrans}_3^{64}(\\
  &\text{ReLU}(\text{ConvTrans}_3^{128}(z_{reshape}))))\big]\big],
\end{aligned}
\end{equation}
where we use sigmoid activation for output because the value range of the observation vector is between 0 and 1. Thus, The loss function is expressed as:
\begin{eqnarray}\label{vaeloss}
  L_{vae}=L_{recon}+L_{kl},
\end{eqnarray}
where $L_{recon}$ and $L_{kl}$ are simply expressed as
\begin{equation}
\begin{aligned}
    L_{recon}&=-\log p(x|z), \\
    L_{kl}&=-\frac{1}{2} \sum \left(1+\log \sigma^2 - \mu^2 - \sigma^2\right).
\end{aligned}
\end{equation}

In short, the VAE can be trained online during the RL training process. Due to the efficient calculation of CNN on GPU, the overall algorithm can maintain its advantage in saving time. The global feature representation $z$ will be concatenated with the local feature to be local-global aggregation representations $s^f_t$ and passed to the corresponding agent for PPO learning.

\subsection{PPO}
We adopt the PPO algorithm with Generalized Advantage Estimation (GAE) trick as backbone model, refer to bottom right of Fig. \ref{fig:total_framework}. The core idea of PPO is to update the policy by maximizing a clipped objective function, which helps prevent large updates that could destabilize training.

The policy function for PPO can be expressed as:
\begin{equation}\label{policyloss}
L^{CLIP}(\theta) = \mathbb{E}_t \left[ \min \left( r_t(\theta) \hat{A}_t, \text{clip}(r_t(\theta), 1 - \epsilon, 1 + \epsilon) \hat{A}_t \right) \right],
\end{equation}
where \( r_t(\theta) \) is the probability ratio defined as \( \frac{\pi_\theta(a_t | o_t)}{\pi_{\theta_{\text{old}}}(a_t | o_t)} \). Here, \( \pi_\theta \) denotes the policy parameterized by \( \theta \), \( a_t \) is the action taken, and \( o_t \) is the local observation at time \( t \). The term \( \hat{A}_t \) represents the estimated advantage, which quantifies how much better the taken action was compared to the expected action under the current policy.

We use GAE to compute the advantage estimate \( \hat{A}_t \). It considers not only the immediate reward but also the value of future states, allowing for a more accurate approximation of advantage. The advantage can be computed as follows:
\begin{equation}\label{GAE}
  \hat{A}_t=\sum_{l=0}^\infty(\gamma\lambda)^l\delta_{t+l},
\end{equation}
where $lambda$ is a discount factor to balance short-term and long-term advantages, and \( \delta_t \) is defined as:
\begin{equation}
\delta_t = r_t + \gamma V_\theta(s^f_{t+1}) - V_\theta(s^f_t),
\end{equation}
where \( V_\theta(s^f) \) represents the value function approximated by the neural network, \( r_t \) is the immediate reward, \(s^f_t\) is global-local aggregation representation introduced in the previous subsection, and \( \gamma \) is the discount factor that balances the importance of future rewards.

In addition, to updating the policy, the value function loss can be defined as:
\begin{equation}\label{valueloss}
L^V(\theta) = \mathbb{E}_t \left[ (V_\theta(s^f_t) - V_{\text{target},t})^2 \right],
\end{equation}
where \( V_{\text{target},t} \) is typically the sum of the immediate reward and the discounted value of the next state:
\begin{equation}
V_{\text{target},t} = r_t + \gamma V_\theta(s^f_{t+1}).
\end{equation}

Through this structured approach, PPO with GAE provides a robust mechanism for policy updates while maintaining stability in learning, allowing for effective exploration and improved sample efficiency in the task. MacLight pseudocode is summarized in Algorithm \ref{alg:MacLight}. Ultimately, the algorithm will try to maximize the total reward to achieve the overall goal.
{\small
\begin{algorithm}[h]
  \caption{The pseudocode of MacLight}\label{alg:MacLight}
  \begin{algorithmic}[1]
  \Ensure {The neural networks: $f^e, f^d, V^k; \pi^k$ // Encoder, Deconder, ValueNet, PolicyNet;}
  \State \textbf{Initialize:} $L, T, K, E$; // Training episodes, timesteps, number of intersections (agents), inner updating epoch of PPO;
  \For{episode $l = 1$ to $L$}
      \For{timestep $t = 1$ to $T$}
          \State  Global feature matrix $s_t$;
          \State  Encoder global representation $s^g_t = f^e(s_t)$;
          \State  Decoder reconstruction $s^r_t = f^d(s^g_t)$;
          \State Update the autoencoder $f^e, f^d$ using Eq. \ref{vaeloss};
          \For{agent $k = 1$ to $K$}
              \State Partial observation $o_t^{k}$, global representation $s^g_t$;
              \State Global-local representation $s^f_t = [s^g_t, o_t^{k}]$;
              \State Calculate advantage $\hat{A}_t$ using $V$, $s^f_t$ by Eq. \ref{GAE};
              \For{train epoch $e = 1$ to $E$}
                \State Update $V^k$ using $s^f_t$ by Eq. \ref{valueloss};
                \State Update $\pi^k$ using $o_t^{k}$ and $\hat{A}_t$ by Eq. \ref{policyloss};
              \EndFor
          \EndFor
      \EndFor
  \EndFor
  \end{algorithmic}
\end{algorithm}
}

\begin{figure}[htbp]
  \centering
  \includegraphics[width=0.95\linewidth]{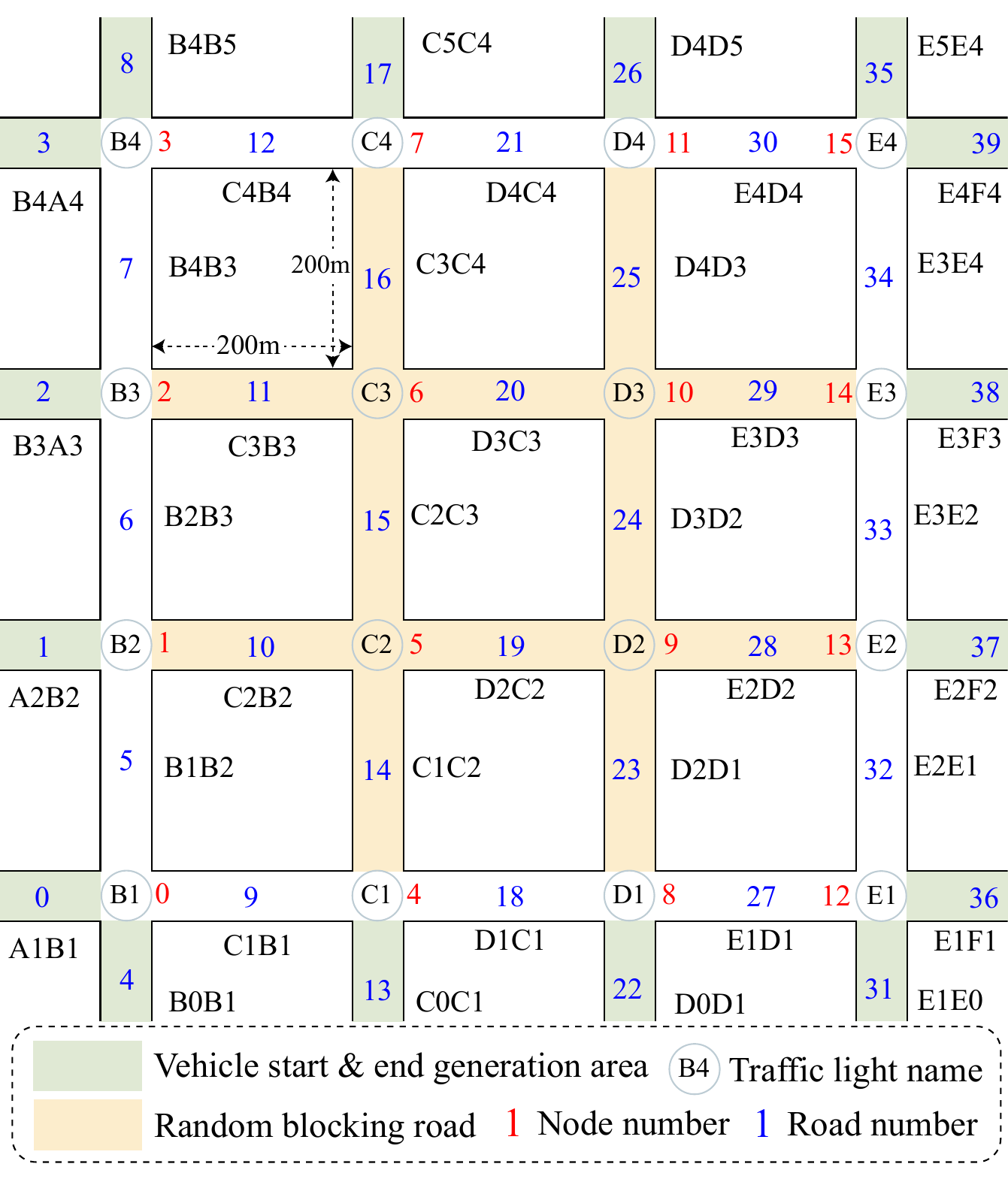}
  \caption{The road network of the simulation environment}
  \label{fig:map_indicator}
  \Description{Road traffic network.}
\end{figure}

\begin{figure}[htbp]
  \centering
  \includegraphics[width=0.9\linewidth]{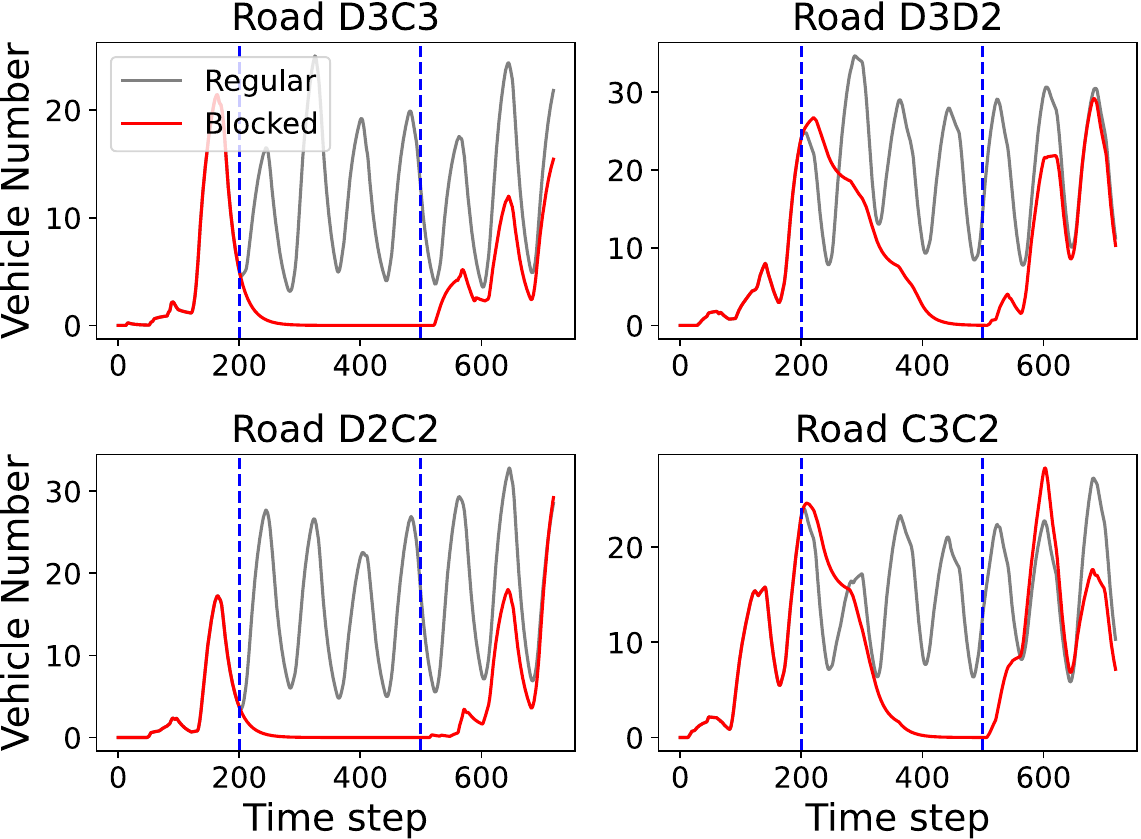}
  \caption{Some blocked lane traffic flow distributions}
  \label{fig:blocked_demo}
  \Description{The blocked road traffic flow distributions.}
\end{figure}

\subsection{Dynamic traffic flow construction}

In current TSC RL studies, the deployment of traffic flow is typically fixed, with all vehicles following predetermined routes. However, our experiment is the first to build a dynamic traffic flow environment that simulates emergency road events, which cause sudden changes in the distribution of traffic on other roads.

\begin{figure}[t]
  \centering
  \includegraphics[width=0.9\linewidth]{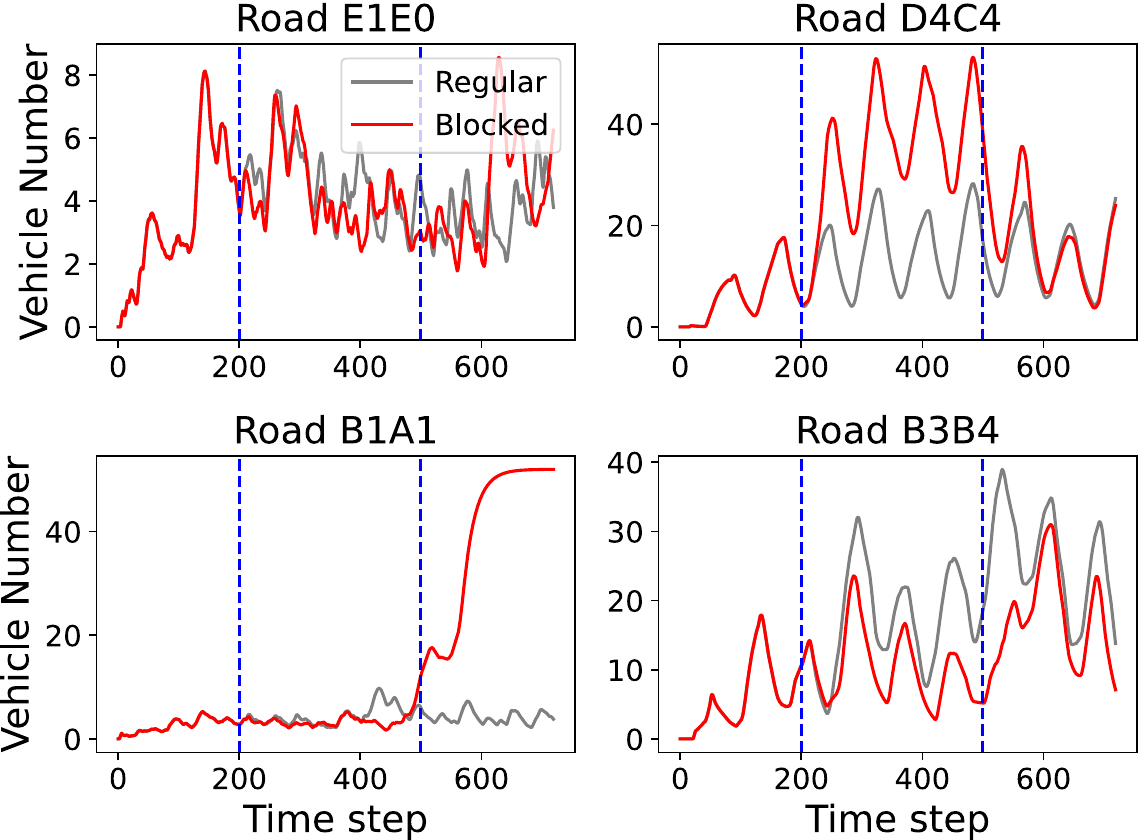}
  \caption{Some regular road traffic flow distributions}
  \label{fig:regular_demo}
  \Description{Other regular road traffic flow distributions.}
\end{figure}

As illustrated in Fig. \ref{fig:map_indicator}, when four central roads—D3C3, D3D2, D2C2 and C3C2—are blocked, all vehicles are asked to reroute. We conduct two experiments to analyze the effects. In Fig. \ref{fig:blocked_demo}, the gray curve shows the traffic flow distribution on a specific lane under normal conditions (without interference), while the red curve shows the distribution after blocking a specific road. The lane blockage period is marked by a blue vertical dashed line. During this time, the traffic volume on the affected lanes drops significantly as vehicles select new optimal routes. Fig. \ref{fig:regular_demo} illustrates the changes in traffic flow on other unblocked roads. Detailed distribution statistics can be found in Appendix \ref{appendix:A}.

The impact of congestion on one road can be quite complex and influence other roads in unpredictable ways. Fig. \ref{fig:regular_demo} shows the flow distribution for unblocked lanes after designated roads are closed. For example, while the traffic flow distribution on roads E1E10 remains largely unaffected, there is a significant increase in traffic on D4C4. On the other hand, B1A1 sees a sharp rise in traffic, while B3B4 experiences a decrease. Such interdependencies are difficult to model accurately but are common in real-world traffic systems. Consequently, our algorithm must account for these complex interactions.

\section{EXPERIMENTS}
This section introduces the experimental environments, evaluation indicators, comparison algorithms, main experimental and ablation analysis. Table \ref{table:main_result} can quickly check the experimental results.

\subsection{Environment and metrics}

\begin{figure}[t]
	\centering
	\subfigure[Average waiting time]{
		\includegraphics[width=0.3\linewidth]{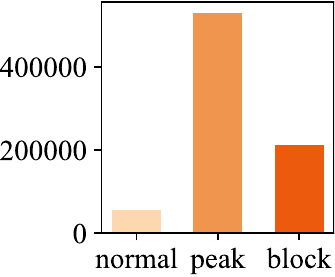}
		\label{}
	}
	\subfigure[Average queue length]{
		\includegraphics[width=0.3\linewidth]{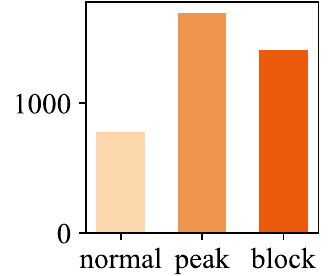}
		\label{}
	}
  \subfigure[Average speed]{
		\includegraphics[width=0.3\linewidth]{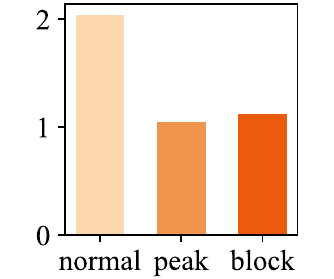}
		\label{}
	}
	\caption{Statistics of different experimental scenarios} \label{fig:scenario_indicators}
  \Description{Statistics of different experimental scenarios}
\end{figure}

\textbf{Environment.} In order to comprehensively evaluate algorithms, three different traffic scenarios are constructed on the same road network as shown in Fig. \ref{fig:map_indicator}. The system is represented by a 4$\times$4 grid arranged horizontally and vertically, with a distance of 200 meters. The three scenarios are a normal-pressure scenario with regular traffic flow called \textbf{Normal}, a high-pressure scenario with extremely high traffic flow called \textbf{Peak}, and a dynamic traffic scenario with normal traffic flow but random emergency road blockage called \textbf{Block}. The randomly blocked lanes are indicated by yellow parts in Fig. \ref{fig:map_indicator}. In terms of task difficulty, the minimum traffic pressure for our scenarios is much greater than all the other current studies, refer to Table \ref{table:flow_compare}. Benchmarking our experimental scenarios, when all three scenarios adopt a fixed phase switching time of every 45 seconds, the system simulation statistics are shown in Fig. \ref{fig:scenario_indicators}. All simulations are performed on the SUMO \cite{SUMO2018} simulation platform.

\textbf{Metrics.} In each scenario, we not only present the total reward results for all algorithms but also establish three objective metrics for comprehensive evaluation: the system's average waiting time, the queue length of waiting vehicles, and the average speed. These metrics take into account both temporal and spatial factors, enabling a more holistic assessment of the transportation system and preventing reward hacking \cite{10.5555/3600270.3600957}. The training seed range for all algorithms is set from 42 to 46, with details provided in the following subsection.

\begin{table}[t]
  \caption{Number of vehicles deployed on different scenario. Normal\&Block and Peak are ours, while arterial4x4 and grid4x4 are the two similar scenarios tested in DuaLight.}\label{table:flow_compare}
  {\small\begin{tabular}{lllll}
  \toprule
            & Normal\&Block & Peak  & arterial4x4 & grid4x4 \\
  \midrule
  Vehicles & \multicolumn{1}{c}{8000}   & \multicolumn{1}{c}{10286}  & \multicolumn{1}{c}{2485} & \multicolumn{1}{c}{1472}  \\
  \bottomrule
  \end{tabular}}
\end{table}

  \begin{table*}[t]
  \caption{Experimental results of each scenario and indicator. IPPO can be considered as an ablation experiment. The specific values in the table include the mean of the current column indicator and the standard deviation in brackets. Best results
  in boldface, and the second-best results underlined. The preferred direction of the indicator is marked by up and down arrows. The waiting time is not given corresponding standard deviation due to the large value.}\label{table:main_result}
  {\small\begin{tabular}{lllllllllllll}
  \toprule
  Scenario  & \multicolumn{4}{c}{Normal}                                                                                        & \multicolumn{4}{c}{Peak}                                                                                          & \multicolumn{4}{c}{Block}                                                                                         \\
  \cmidrule(lr){1-1} \cmidrule(lr){2-5} \cmidrule(lr){6-9} \cmidrule(lr){10-13}
  Indicator & \multicolumn{1}{c}{Return↑} & \multicolumn{1}{c}{Wait↓} & \multicolumn{1}{c}{Queue↓} & \multicolumn{1}{c}{Speed↑} & \multicolumn{1}{c}{Return↑} & \multicolumn{1}{c}{Wait↓} & \multicolumn{1}{c}{Queue↓} & \multicolumn{1}{c}{Speed↑} & \multicolumn{1}{c}{Return↑} & \multicolumn{1}{c}{Wait↓} & \multicolumn{1}{c}{Queue↓} & \multicolumn{1}{c}{Speed↑} \\
  \midrule
  Fixed time     & -37.14\tiny{(9)}                & 56409              & 785\tiny{(170)}                   & 2.1\tiny{(1)}                   & \textbf{-171.0}\tiny{(98)}       & \textbf{292582}    & 1526\tiny{(236)}                  & 1.2\tiny{(1)}                   & -179.5\tiny{(59.3)}              & 253944           & 1477\tiny{(210)}                  & 1.1\tiny{(0)}                   \\
  IPPO      & \underline{-6.6}\tiny{(22)}     & \underline{10254}  & \underline{152}\tiny{(114)}       & \underline{6.2}\tiny{(1)}       & -434.8\tiny{(451)}               & 1258399            & \underline{1456}\tiny{(1262)}     & \underline{2.8}\tiny{(3)}       & \textbf{-12.0}\tiny{(28)}        & \textbf{13144}   & \textbf{221}\tiny{(197)}          & \textbf{5.4}\tiny{(1)}          \\
  MAPPO     & -67.8\tiny{(65)}                & 122793             & 509\tiny{(186)}                   & 3.3\tiny{(1)}                   & -924.2\tiny{(117)}               & 2559550            & 2942\tiny{(165)}                  & 0.1\tiny{(0)}                   & -127.8\tiny{(81)}                & 190998           & 972\tiny{(211)}                   & 1.7\tiny{(0)}                   \\
  IDQN      & -598.2\tiny{(276)}              & 1650798            & 2474\tiny{(956)}                  & 0.7\tiny{(1)}                   & -1054.4\tiny{(101)}              & 4546469            & 3498\tiny{(248)}                  & 0.0\tiny{(0)}                   & -796.6\tiny{(198)}               & 2625699          & 3136\tiny{(610)}                  & 0.1\tiny{(0)}                   \\
  CoLight   & -716.2\tiny{(283)}              & 2538938            & 2913\tiny{(969)}                  & 0.5\tiny{(1)}                   & -969.5\tiny{(146)}               & 4747313            & 3438\tiny{(436)}                  & 0.0\tiny{(0)}                   & -788.1\tiny{(228)}               & 3360669          & 3186\tiny{(772)}                  & 0.2\tiny{(1)}                   \\
  DuaLight  & -712.8\tiny{(293)}              & 2630974            & 2858\tiny{(1009)}                 & 0.5\tiny{(1)}                   & -977.9\tiny{(154)}               & 4664564            & 3410\tiny{(423)}                  & 0.0\tiny{(0)}                   & -770.5\tiny{(246)}               & 3221476          & 3146\tiny{(816)}                  & 0.2\tiny{(1)}                   \\
  \midrule
  MacLight  & \textbf{-4.02}\tiny{(10)}       & \textbf{4737}      & \textbf{140}\tiny{(90)}           & \textbf{6.3}\tiny{(1)}          & \underline{-362.3}\tiny{(423)}   & \underline{998411} & \textbf{1267}\tiny{(1237)}        & \textbf{3.3}\tiny{(3)}          & \underline{-17.3}\tiny{(44.0)}   & \underline{24224}& \underline{249}\tiny{(237)}       & \underline{5.2}\tiny{(1)}       \\
  \bottomrule
  \end{tabular}}
\end{table*}

\begin{figure*}[h]
	\centering
	\subfigure[Normal]{
		\includegraphics[width=0.31\linewidth]{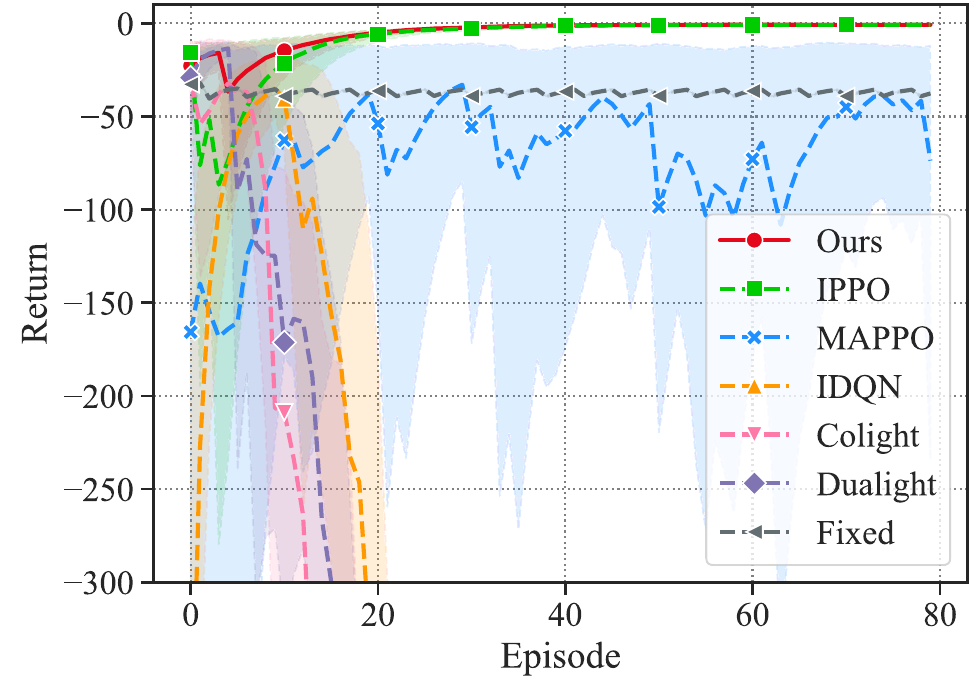}
		\label{}
	}
	\subfigure[Peak]{
		\includegraphics[width=0.31\linewidth]{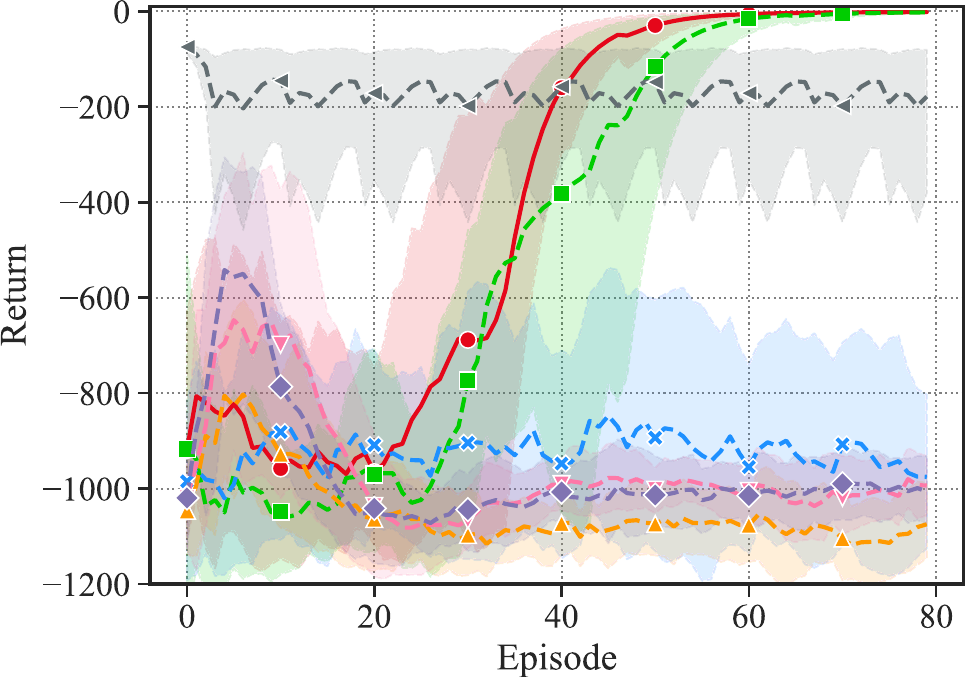}
		\label{}
	}
  \subfigure[Block]{
		\includegraphics[width=0.31\linewidth]{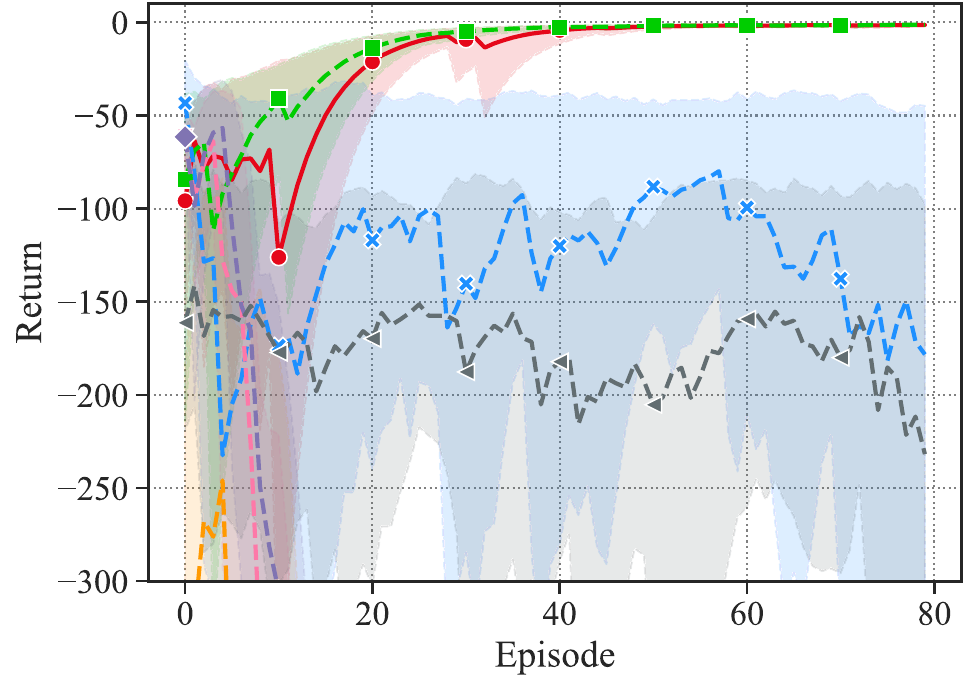}
		\label{}
	}
	\caption{Training details of cumulative rewards} \label{fig:training_process}
  \Description{Training details of cumulative rewards.}
\end{figure*}

\subsection{Comparison methods}

MacLight will be compared with a variety of algorithms, including the traditional method of setting a fixed time switching phase and a variety of advanced algorithms based on RL. MacLight's model parameters refer to Appendix \ref{appendix:B}.

\textbf{Fixed time.} Similar to the control method in reality, we configure the same fixed time switching method for all traffic lights: switching the phase every 45 seconds.

\textbf{IPPO.} Refer to \cite{IPPO}. A separate agent with PPO algorithm is constructed for each intersection, and each agent only focuses on its own local information. IPPO can be regarded as the ablation object of MacLight.

\textbf{MAPPO.} Refer to \cite{MAPPO}. Similar to IPPO, but only one value evaluation network is used globally, whose input is the concatenation of local observations of all agents, while policy modules are as same as IPPO.

\textbf{IDQN.} Similar to IPPO, but replaces the PPO with DQN. IDQN is the backbone model of CoLight and DuaLight.

\textbf{CoLight.} Referring to \cite{colight}, a strong algorithm for applying RL to TSC tasks using GAT, built on top of DQN.

\textbf{DuaLight.} Reference \cite{DuaLight}, a SOTA based on CoLight, adds feature weight matrix and neighborhood weight matrix for different scenarios to the backbone network for Q learning, which shows better representation effect than CoLight. It is also based on DQN.

\subsection{Main results}

\textbf{Comparative experiments.} Table \ref{table:main_result} shows comprehensive comparison of experimental results. MacLight performs best in the Normal scene, with relatively good average performance and stability, followed by IPPO. In the high-pressure traffic environment Peak, the return and waiting time indicators are not as good as the Fixed method, because the indicators represent the average of the entire process, and if we check the final value, MacLight still has the best performance. In the dynamic traffic environment Block, indicators are inferior to IPPO. IDQN and the DQN-based CoLight and DuaLight methods perform poorly and are very prone to overfitting and policy collapse when faced with relatively sparse rewards and unstable data.

\textbf{Training and testing.} Fig. \ref{fig:training_process} shows the change of cumulative rewards during the entire training process, with the shadows indicating the maximum and minimum regions recorded for different seed experiments. On-policy approaches MacLight and IPPO, consistently demonstrates stable policy improvement across all scenarios. In contrast, off-policy methods such as IDQN and CoLight, while exhibiting robust initial performance, tend to collapse shortly thereafter. These methods are better suited for less challenging scenarios, leveraging the advantages of smaller models to avoid overfitting. However, they falter in high-difficulty, sparse-reward environments. For complete training diagrams of all indicators, refer to Appendix \ref{appendix:C}. We show the test results of all algorithms in Table. \ref{table:test}, where the indicator is average return.

\textbf{Ablation analysis.} IPPO in Table \ref{table:main_result} can be regarded as an ablation experiment of MacLight, because MacLight modifies the input of the value module from local features to local-global aggregate representation. On most indicators, MacLight shows advantages, while the second-best method is IPPO.

\textbf{Training time on wall clock.} Table \ref{table:training_time} shows the training time of MacLight compared to other off-policy algorithms. We tested multiple random number seeds, each seed trained 80 episodes, and each episode contained 3600 seconds simulation. The times in the table are calculated as the average of the total length of 80 episodes on each seed. MacLight requires less than 1 hour to train, while off-policy algorithm IDQN needs at least 2 hours, Colight and DuaLight are even slower. This is because the GCN-based method cannot be massively parallelized, further slowing down the computational efficiency.

\begin{table}[t]
  \caption{Test results of each algorithm on the average return. Slight differences from the training metrics can be noticed.}\label{table:test}
  \centering
  {\small\begin{tabular}{llll}
  \toprule
      ~          & Normal         & Peak     & Block \\ \midrule
      Fixed      & -31            & -100     & -312 \\
      IPPO       & \textbf{-0.68} & -1.60    & -1.18 \\
      MAPPO      & -202           & -977     & -206 \\
      IDQN       & -842           & -1051    & -891 \\
      CoLight    & -937           & -997     & -942 \\
      DuaLight   & -878           & -1034    & -876 \\ \midrule
      MacLight   & -0.71          & \textbf{-1.46}    & \textbf{-1.17} \\ \bottomrule
  \end{tabular}}
\end{table}

\begin{table}[t]
  \caption{Comparison of training wall time (minute) for 80 episodes between MacLight (Ours) and off-policy methods. All algorithms run on a single A100. }\label{table:training_time}
  {\small\begin{tabular}{lllll}
  \toprule
            & Normal & Peak  & Block\\
  \midrule
  IDQN     & 137.4  & 178.9 & 186.4\\
  CoLight  & 373.7  & 405.1 & 391.7\\
  DuaLight & 413.2  & 456.4 & 283.2\\
  \midrule
  MacLight & 43.0   & 58.1  & 39.1 \\
  \bottomrule
  \end{tabular}}
\end{table}


\section{CONCLUSION}

In this paper, we proposed the MacLight for TSC and construct both static and dynamic traffic flow for evaluation. The main contribution of MacLight is to construct a CNN-based VAE for global state feature extraction, and connect with the local state to form a local-global representation, which is used as the input of the value evaluation module to guide the policy improvement. MacLight uses the PPO algorithm as the backbone so that global and local information can be processed in parallel and improve each other. In addition, as an on-policy algorithm, MacLight provides high operating efficiency, taking only about one-third of the time of the off-policy method. Finally, the dynamic traffic simulation environment we constructed greatly expands the current research space and provides a basis for applying RL in emergency traffic scenarios.

There is still room for improvement in our work. Although CNN is more efficient than GCN-based methods, real road networks are usually not as regular as Manhattan roads and cannot be directly constructed as pixel matrices. We can use multiscale convolution to alleviate this problem in the future. In addition, there is room for improvement in the reward function, such as introducing both local and global indicators.

\balance
\begin{acks}
This work was supported by Hubei Provincial Key
Laboratory of Metallurgical Industry Process System Science (Y202105) and High Performance Computing Center of Wuhan University of
Science and Technology. We thank the reviewers for their contributions.
\end{acks}


\bibliographystyle{ACM-Reference-Format}
\bibliography{references}


\begin{thebibliography}{32}


\ifx \showCODEN    \undefined \def \showCODEN     #1{\unskip}     \fi
\ifx \showDOI      \undefined \def \showDOI       #1{#1}\fi
\ifx \showISBNx    \undefined \def \showISBNx     #1{\unskip}     \fi
\ifx \showISBNxiii \undefined \def \showISBNxiii  #1{\unskip}     \fi
\ifx \showISSN     \undefined \def \showISSN      #1{\unskip}     \fi
\ifx \showLCCN     \undefined \def \showLCCN      #1{\unskip}     \fi
\ifx \shownote     \undefined \def \shownote      #1{#1}          \fi
\ifx \showarticletitle \undefined \def \showarticletitle #1{#1}   \fi
\ifx \showURL      \undefined \def \showURL       {\relax}        \fi
\providecommand\bibfield[2]{#2}
\providecommand\bibinfo[2]{#2}
\providecommand\natexlab[1]{#1}
\providecommand\showeprint[2][]{arXiv:#2}

\bibitem[Alegre(2019)]%
        {sumorl}
\bibfield{author}{\bibinfo{person}{Lucas~N. Alegre}.} \bibinfo{year}{2019}\natexlab{}.
\newblock \bibinfo{title}{{SUMO-RL}}.
\newblock \bibinfo{howpublished}{\url{https://github.com/LucasAlegre/sumo-rl}}.
\newblock


\bibitem[Alegre et~al\mbox{.}(2021)]%
        {Alegre2021}
\bibfield{author}{\bibinfo{person}{Lucas~N. Alegre}, \bibinfo{person}{Ana L.~C. Bazzan}, {and} \bibinfo{person}{Bruno~C. da Silva}.} \bibinfo{year}{2021}\natexlab{}.
\newblock \showarticletitle{Quantifying the impact of non-stationarity in reinforcement learning-based traffic signal control}.
\newblock \bibinfo{journal}{\emph{PeerJ Computer Science}}  \bibinfo{volume}{7} (\bibinfo{year}{2021}), \bibinfo{pages}{e575}.
\newblock
\urldef\tempurl%
\url{https://doi.org/10.7717/peerj-cs.575}
\showDOI{\tempurl}


\bibitem[Ault and Sharon(2021)]%
        {ault2021reinforcement}
\bibfield{author}{\bibinfo{person}{James Ault} {and} \bibinfo{person}{Guni Sharon}.} \bibinfo{year}{2021}\natexlab{}.
\newblock \showarticletitle{Reinforcement Learning Benchmarks for Traffic Signal Control}. In \bibinfo{booktitle}{\emph{Proceedings of the Thirty-fifth Conference on Neural Information Processing Systems (NeurIPS 2021) Datasets and Benchmarks Track}}.
\newblock


\bibitem[Chen et~al\mbox{.}(2020)]%
        {Thousand}
\bibfield{author}{\bibinfo{person}{Chacha Chen}, \bibinfo{person}{Hua Wei}, \bibinfo{person}{Nan Xu}, \bibinfo{person}{Guanjie Zheng}, \bibinfo{person}{Ming Yang}, \bibinfo{person}{Yuanhao Xiong}, \bibinfo{person}{Kai Xu}, {and} \bibinfo{person}{Zhenhui Li}.} \bibinfo{year}{2020}\natexlab{}.
\newblock \showarticletitle{Toward A Thousand Lights: Decentralized Deep Reinforcement Learning for Large-Scale Traffic Signal Control}.
\newblock \bibinfo{journal}{\emph{Proceedings of the AAAI Conference on Artificial Intelligence}} \bibinfo{volume}{34}, \bibinfo{number}{04} (\bibinfo{date}{Apr.} \bibinfo{year}{2020}), \bibinfo{pages}{3414--3421}.
\newblock
\urldef\tempurl%
\url{https://doi.org/10.1609/aaai.v34i04.5744}
\showDOI{\tempurl}


\bibitem[de~Witt et~al\mbox{.}(2020)]%
        {IPPO}
\bibfield{author}{\bibinfo{person}{Christian~Schroeder de Witt}, \bibinfo{person}{Tarun Gupta}, \bibinfo{person}{Denys Makoviichuk}, \bibinfo{person}{Viktor Makoviychuk}, \bibinfo{person}{Philip H.~S. Torr}, \bibinfo{person}{Mingfei Sun}, {and} \bibinfo{person}{Shimon Whiteson}.} \bibinfo{year}{2020}\natexlab{}.
\newblock \bibinfo{title}{Is Independent Learning All You Need in the StarCraft Multi-Agent Challenge?}
\newblock
\newblock
\showeprint[arxiv]{2011.09533}~[cs.AI]
\urldef\tempurl%
\url{https://arxiv.org/abs/2011.09533}
\showURL{%
\tempurl}


\bibitem[Diakaki et~al\mbox{.}(2002)]%
        {DIAKAKI2002183}
\bibfield{author}{\bibinfo{person}{Christina Diakaki}, \bibinfo{person}{Markos Papageorgiou}, {and} \bibinfo{person}{Kostas Aboudolas}.} \bibinfo{year}{2002}\natexlab{}.
\newblock \showarticletitle{A multivariable regulator approach to traffic-responsive network-wide signal control}.
\newblock \bibinfo{journal}{\emph{Control Engineering Practice}} \bibinfo{volume}{10}, \bibinfo{number}{2} (\bibinfo{year}{2002}), \bibinfo{pages}{183--195}.
\newblock
\showISSN{0967-0661}
\urldef\tempurl%
\url{https://doi.org/10.1016/S0967-0661(01)00121-6}
\showDOI{\tempurl}


\bibitem[Gershenson(2005)]%
        {gershenson2005selforganizingtrafficlights}
\bibfield{author}{\bibinfo{person}{Carlos Gershenson}.} \bibinfo{year}{2005}\natexlab{}.
\newblock \bibinfo{title}{Self-Organizing Traffic Lights}.
\newblock
\newblock
\showeprint[arxiv]{nlin/0411066}~[nlin.AO]
\urldef\tempurl%
\url{https://arxiv.org/abs/nlin/0411066}
\showURL{%
\tempurl}


\bibitem[Hochreiter and Schmidhuber(1997)]%
        {LSTM}
\bibfield{author}{\bibinfo{person}{Sepp Hochreiter} {and} \bibinfo{person}{J\"{u}rgen Schmidhuber}.} \bibinfo{year}{1997}\natexlab{}.
\newblock \showarticletitle{Long Short-Term Memory}.
\newblock \bibinfo{journal}{\emph{Neural Comput.}} \bibinfo{volume}{9}, \bibinfo{number}{8} (\bibinfo{date}{Nov.} \bibinfo{year}{1997}), \bibinfo{pages}{1735–1780}.
\newblock
\showISSN{0899-7667}
\urldef\tempurl%
\url{https://doi.org/10.1162/neco.1997.9.8.1735}
\showDOI{\tempurl}


\bibitem[Jiang et~al\mbox{.}(2024)]%
        {GuideLight}
\bibfield{author}{\bibinfo{person}{Haoyuan Jiang}, \bibinfo{person}{Xuantang Xiong}, \bibinfo{person}{Ziyue Li}, \bibinfo{person}{Hangyu Mao}, \bibinfo{person}{Guanghu Sui}, \bibinfo{person}{Jingqing Ruan}, \bibinfo{person}{Yuheng Cheng}, \bibinfo{person}{Hua Wei}, \bibinfo{person}{Wolfgang Ketter}, {and} \bibinfo{person}{Rui Zhao}.} \bibinfo{year}{2024}\natexlab{}.
\newblock \bibinfo{title}{GuideLight: "Industrial Solution" Guidance for More Practical Traffic Signal Control Agents}.
\newblock
\newblock
\showeprint[arxiv]{2407.10811}~[cs.MA]
\urldef\tempurl%
\url{https://arxiv.org/abs/2407.10811}
\showURL{%
\tempurl}


\bibitem[Kingma and Welling(2014)]%
        {VAE}
\bibfield{author}{\bibinfo{person}{Diederik~P. Kingma} {and} \bibinfo{person}{Max Welling}.} \bibinfo{year}{2014}\natexlab{}.
\newblock \showarticletitle{Auto-Encoding Variational Bayes}. In \bibinfo{booktitle}{\emph{2nd International Conference on Learning Representations, {ICLR} 2014, Banff, AB, Canada, April 14-16, 2014, Conference Track Proceedings}}, \bibfield{editor}{\bibinfo{person}{Yoshua Bengio} {and} \bibinfo{person}{Yann LeCun}} (Eds.).
\newblock
\urldef\tempurl%
\url{http://arxiv.org/abs/1312.6114}
\showURL{%
\tempurl}


\bibitem[Krizhevsky et~al\mbox{.}(2012)]%
        {NIPS2012c399862d}
\bibfield{author}{\bibinfo{person}{Alex Krizhevsky}, \bibinfo{person}{Ilya Sutskever}, {and} \bibinfo{person}{Geoffrey~E Hinton}.} \bibinfo{year}{2012}\natexlab{}.
\newblock \showarticletitle{ImageNet Classification with Deep Convolutional Neural Networks}. In \bibinfo{booktitle}{\emph{Advances in Neural Information Processing Systems}}, \bibfield{editor}{\bibinfo{person}{F.~Pereira}, \bibinfo{person}{C.J. Burges}, \bibinfo{person}{L.~Bottou}, {and} \bibinfo{person}{K.Q. Weinberger}} (Eds.), Vol.~\bibinfo{volume}{25}. \bibinfo{publisher}{Curran Associates, Inc.}
\newblock


\bibitem[Lea et~al\mbox{.}(2016)]%
        {TCN}
\bibfield{author}{\bibinfo{person}{Colin Lea}, \bibinfo{person}{Ren{\'{e}} Vidal}, \bibinfo{person}{Austin Reiter}, {and} \bibinfo{person}{Gregory~D. Hager}.} \bibinfo{year}{2016}\natexlab{}.
\newblock \showarticletitle{Temporal Convolutional Networks: A Unified Approach to Action Segmentation}.
\newblock \bibinfo{journal}{\emph{CoRR}}  \bibinfo{volume}{abs/1608.08242} (\bibinfo{year}{2016}).
\newblock
\showeprint[arXiv]{1608.08242}


\bibitem[Lopez et~al\mbox{.}(2018)]%
        {SUMO2018}
\bibfield{author}{\bibinfo{person}{Pablo~Alvarez Lopez}, \bibinfo{person}{Michael Behrisch}, \bibinfo{person}{Laura Bieker-Walz}, \bibinfo{person}{Jakob Erdmann}, \bibinfo{person}{Yun-Pang Fl{\"o}tter{\"o}d}, \bibinfo{person}{Robert Hilbrich}, \bibinfo{person}{Leonhard L{\"u}cken}, \bibinfo{person}{Johannes Rummel}, \bibinfo{person}{Peter Wagner}, {and} \bibinfo{person}{Evamarie Wie{\ss}ner}.} \bibinfo{year}{2018}\natexlab{}.
\newblock \showarticletitle{Microscopic Traffic Simulation using SUMO}, In \bibinfo{booktitle}{The 21st IEEE International Conference on Intelligent Transportation Systems}.
\newblock \bibinfo{journal}{\emph{IEEE Intelligent Transportation Systems Conference (ITSC)}}.
\newblock


\bibitem[Lou et~al\mbox{.}(2022)]%
        {Metalight}
\bibfield{author}{\bibinfo{person}{Yican Lou}, \bibinfo{person}{Jia Wu}, {and} \bibinfo{person}{Yunchuan Ran}.} \bibinfo{year}{2022}\natexlab{}.
\newblock \showarticletitle{Meta-Reinforcement Learning for Multiple Traffic Signals Control}. In \bibinfo{booktitle}{\emph{Proceedings of the 31st ACM International Conference on Information \& Knowledge Management}} (Atlanta, GA, USA) \emph{(\bibinfo{series}{CIKM '22})}. \bibinfo{publisher}{Association for Computing Machinery}, \bibinfo{address}{New York, NY, USA}, \bibinfo{pages}{4264–4268}.
\newblock
\showISBNx{9781450392365}
\urldef\tempurl%
\url{https://doi.org/10.1145/3511808.3557640}
\showDOI{\tempurl}


\bibitem[Lu et~al\mbox{.}(2024)]%
        {DuaLight}
\bibfield{author}{\bibinfo{person}{Jiaming Lu}, \bibinfo{person}{Jingqing Ruan}, \bibinfo{person}{Haoyuan Jiang}, \bibinfo{person}{Ziyue Li}, \bibinfo{person}{Hangyu Mao}, {and} \bibinfo{person}{Rui Zhao}.} \bibinfo{year}{2024}\natexlab{}.
\newblock \showarticletitle{DuaLight: Enhancing Traffic Signal Control by Leveraging Scenario-Specific and Scenario-Shared Knowledge}. In \bibinfo{booktitle}{\emph{Proceedings of the 23rd International Conference on Autonomous Agents and Multiagent Systems}} (Auckland, New Zealand) \emph{(\bibinfo{series}{AAMAS '24})}. \bibinfo{publisher}{International Foundation for Autonomous Agents and Multiagent Systems}, \bibinfo{address}{Richland, SC}, \bibinfo{pages}{1283–1291}.
\newblock
\showISBNx{9798400704864}


\bibitem[Ma and Wu(2020)]%
        {SUMOearly}
\bibfield{author}{\bibinfo{person}{Jinming Ma} {and} \bibinfo{person}{Feng Wu}.} \bibinfo{year}{2020}\natexlab{}.
\newblock \showarticletitle{Feudal Multi-Agent Deep Reinforcement Learning for Traffic Signal Control}. In \bibinfo{booktitle}{\emph{Proceedings of the 19th International Conference on Autonomous Agents and MultiAgent Systems}} (Auckland, New Zealand) \emph{(\bibinfo{series}{AAMAS '20})}. \bibinfo{publisher}{International Foundation for Autonomous Agents and Multiagent Systems}, \bibinfo{address}{Richland, SC}, \bibinfo{pages}{816–824}.
\newblock
\showISBNx{9781450375184}


\bibitem[Mnih et~al\mbox{.}(2013)]%
        {DQN}
\bibfield{author}{\bibinfo{person}{Volodymyr Mnih}, \bibinfo{person}{Koray Kavukcuoglu}, \bibinfo{person}{David Silver}, \bibinfo{person}{Alex Graves}, \bibinfo{person}{Ioannis Antonoglou}, \bibinfo{person}{Daan Wierstra}, {and} \bibinfo{person}{Martin Riedmiller}.} \bibinfo{year}{2013}\natexlab{}.
\newblock \bibinfo{title}{Playing Atari with Deep Reinforcement Learning}.
\newblock
\newblock
\showeprint[arxiv]{1312.5602}~[cs.LG]


\bibitem[Oroojlooy et~al\mbox{.}(2020)]%
        {AttendLight}
\bibfield{author}{\bibinfo{person}{Afshin Oroojlooy}, \bibinfo{person}{Mohammadreza Nazari}, \bibinfo{person}{Davood Hajinezhad}, {and} \bibinfo{person}{Jorge Silva}.} \bibinfo{year}{2020}\natexlab{}.
\newblock \showarticletitle{AttendLight: universal attention-based reinforcement learning model for traffic signal control}. In \bibinfo{booktitle}{\emph{Proceedings of the 34th International Conference on Neural Information Processing Systems}} (Vancouver, BC, Canada) \emph{(\bibinfo{series}{NIPS '20})}. \bibinfo{publisher}{Curran Associates Inc.}, \bibinfo{address}{Red Hook, NY, USA}, Article \bibinfo{articleno}{343}, \bibinfo{numpages}{12}~pages.
\newblock
\showISBNx{9781713829546}


\bibitem[Schulman et~al\mbox{.}(2017)]%
        {PPO}
\bibfield{author}{\bibinfo{person}{John Schulman}, \bibinfo{person}{Filip Wolski}, \bibinfo{person}{Prafulla Dhariwal}, \bibinfo{person}{Alec Radford}, {and} \bibinfo{person}{Oleg Klimov}.} \bibinfo{year}{2017}\natexlab{}.
\newblock \bibinfo{title}{Proximal Policy Optimization Algorithms}.
\newblock
\newblock
\showeprint[arxiv]{1707.06347}~[cs.LG]


\bibitem[Skalse et~al\mbox{.}(2024)]%
        {10.5555/3600270.3600957}
\bibfield{author}{\bibinfo{person}{Joar Skalse}, \bibinfo{person}{Nikolaus H.~R. Howe}, \bibinfo{person}{Dmitrii Krasheninnikov}, {and} \bibinfo{person}{David Krueger}.} \bibinfo{year}{2024}\natexlab{}.
\newblock \showarticletitle{Defining and characterizing reward hacking}. In \bibinfo{booktitle}{\emph{Proceedings of the 36th International Conference on Neural Information Processing Systems}} (New Orleans, LA, USA) \emph{(\bibinfo{series}{NIPS '22})}. \bibinfo{publisher}{Curran Associates Inc.}, \bibinfo{address}{Red Hook, NY, USA}, Article \bibinfo{articleno}{687}, \bibinfo{numpages}{12}~pages.
\newblock
\showISBNx{9781713871088}


\bibitem[Sutton(1988)]%
        {Sutton1988}
\bibfield{author}{\bibinfo{person}{Richard~S. Sutton}.} \bibinfo{year}{1988}\natexlab{}.
\newblock \showarticletitle{Learning to predict by the methods of temporal differences}.
\newblock \bibinfo{journal}{\emph{Machine Learning}} \bibinfo{volume}{3}, \bibinfo{number}{1} (\bibinfo{date}{01 Aug} \bibinfo{year}{1988}), \bibinfo{pages}{9--44}.
\newblock
\showISSN{1573-0565}
\urldef\tempurl%
\url{https://doi.org/10.1007/BF00115009}
\showDOI{\tempurl}


\bibitem[Sutton and Barto(2018)]%
        {sutton2018reinforcement}
\bibfield{author}{\bibinfo{person}{Richard~S Sutton} {and} \bibinfo{person}{Andrew~G Barto}.} \bibinfo{year}{2018}\natexlab{}.
\newblock \bibinfo{booktitle}{\emph{Reinforcement learning: An introduction}}.
\newblock \bibinfo{publisher}{MIT press}.
\newblock


\bibitem[Sutton et~al\mbox{.}(1999)]%
        {sutton1999}
\bibfield{author}{\bibinfo{person}{Richard~S. Sutton}, \bibinfo{person}{David McAllester}, \bibinfo{person}{Satinder Singh}, {and} \bibinfo{person}{Yishay Mansour}.} \bibinfo{year}{1999}\natexlab{}.
\newblock \showarticletitle{Policy Gradient Methods for Reinforcement Learning with Function Approximation}. In \bibinfo{booktitle}{\emph{Proceedings of the 12th International Conference on Neural Information Processing Systems}} (Denver, CO) \emph{(\bibinfo{series}{NIPS'99})}. \bibinfo{publisher}{MIT Press}, \bibinfo{address}{Cambridge, MA, USA}, \bibinfo{pages}{1057–1063}.
\newblock


\bibitem[Varaiya(2013)]%
        {Varaiya2013}
\bibfield{author}{\bibinfo{person}{Pravin Varaiya}.} \bibinfo{year}{2013}\natexlab{}.
\newblock \bibinfo{booktitle}{\emph{The Max-Pressure Controller for Arbitrary Networks of Signalized Intersections}}.
\newblock \bibinfo{publisher}{Springer New York}, \bibinfo{address}{New York, NY}, \bibinfo{pages}{27--66}.
\newblock
\showISBNx{978-1-4614-6243-9}
\urldef\tempurl%
\url{https://doi.org/10.1007/978-1-4614-6243-9_2}
\showDOI{\tempurl}


\bibitem[Veli{\v{c}}kovi{\'{c}} et~al\mbox{.}(2018)]%
        {velickovic2018graph}
\bibfield{author}{\bibinfo{person}{Petar Veli{\v{c}}kovi{\'{c}}}, \bibinfo{person}{Guillem Cucurull}, \bibinfo{person}{Arantxa Casanova}, \bibinfo{person}{Adriana Romero}, \bibinfo{person}{Pietro Li{\`{o}}}, {and} \bibinfo{person}{Yoshua Bengio}.} \bibinfo{year}{2018}\natexlab{}.
\newblock \showarticletitle{{Graph Attention Networks}}.
\newblock \bibinfo{journal}{\emph{International Conference on Learning Representations}} (\bibinfo{year}{2018}).
\newblock


\bibitem[Wang et~al\mbox{.}(2022)]%
        {9240060}
\bibfield{author}{\bibinfo{person}{Y. Wang}, \bibinfo{person}{T. Xu}, \bibinfo{person}{X. Niu}, \bibinfo{person}{C. Tan}, \bibinfo{person}{E. Chen}, {and} \bibinfo{person}{H. Xiong}.} \bibinfo{year}{2022}\natexlab{}.
\newblock \showarticletitle{STMARL: A Spatio-Temporal Multi-Agent Reinforcement Learning Approach for Cooperative Traffic Light Control}.
\newblock \bibinfo{journal}{\emph{IEEE Transactions on Mobile Computing}} \bibinfo{volume}{21}, \bibinfo{number}{06} (\bibinfo{date}{jun} \bibinfo{year}{2022}), \bibinfo{pages}{2228--2242}.
\newblock
\showISSN{1558-0660}
\urldef\tempurl%
\url{https://doi.org/10.1109/TMC.2020.3033782}
\showDOI{\tempurl}


\bibitem[Watkins and Dayan(1992)]%
        {Watkins1992}
\bibfield{author}{\bibinfo{person}{Christopher J. C.~H. Watkins} {and} \bibinfo{person}{Peter Dayan}.} \bibinfo{year}{1992}\natexlab{}.
\newblock \showarticletitle{Q-learning}.
\newblock \bibinfo{journal}{\emph{Machine Learning}} \bibinfo{volume}{8}, \bibinfo{number}{3} (\bibinfo{date}{01 May} \bibinfo{year}{1992}), \bibinfo{pages}{279--292}.
\newblock
\showISSN{1573-0565}
\urldef\tempurl%
\url{https://doi.org/10.1007/BF00992698}
\showDOI{\tempurl}


\bibitem[Wei et~al\mbox{.}(2019a)]%
        {presslight}
\bibfield{author}{\bibinfo{person}{Hua Wei}, \bibinfo{person}{Chacha Chen}, \bibinfo{person}{Guanjie Zheng}, \bibinfo{person}{Kan Wu}, \bibinfo{person}{Vikash Gayah}, \bibinfo{person}{Kai Xu}, {and} \bibinfo{person}{Zhenhui Li}.} \bibinfo{year}{2019}\natexlab{a}.
\newblock \showarticletitle{PressLight: Learning Max Pressure Control to Coordinate Traffic Signals in Arterial Network}. In \bibinfo{booktitle}{\emph{Proceedings of the 25th ACM SIGKDD International Conference on Knowledge Discovery \& Data Mining}} (Anchorage, AK, USA) \emph{(\bibinfo{series}{KDD '19})}. \bibinfo{publisher}{Association for Computing Machinery}, \bibinfo{address}{New York, NY, USA}, \bibinfo{pages}{1290–1298}.
\newblock
\showISBNx{9781450362016}
\urldef\tempurl%
\url{https://doi.org/10.1145/3292500.3330949}
\showDOI{\tempurl}


\bibitem[Wei et~al\mbox{.}(2019b)]%
        {colight}
\bibfield{author}{\bibinfo{person}{Hua Wei}, \bibinfo{person}{Nan Xu}, \bibinfo{person}{Huichu Zhang}, \bibinfo{person}{Guanjie Zheng}, \bibinfo{person}{Xinshi Zang}, \bibinfo{person}{Chacha Chen}, \bibinfo{person}{Weinan Zhang}, \bibinfo{person}{Yanmin Zhu}, \bibinfo{person}{Kai Xu}, {and} \bibinfo{person}{Zhenhui Li}.} \bibinfo{year}{2019}\natexlab{b}.
\newblock \showarticletitle{CoLight: Learning Network-level Cooperation for Traffic Signal Control}. In \bibinfo{booktitle}{\emph{Proceedings of the 28th ACM International Conference on Information and Knowledge Management}} (Beijing, China) \emph{(\bibinfo{series}{CIKM '19})}. \bibinfo{publisher}{Association for Computing Machinery}, \bibinfo{address}{New York, NY, USA}, \bibinfo{pages}{1913–1922}.
\newblock
\showISBNx{9781450369763}
\urldef\tempurl%
\url{https://doi.org/10.1145/3357384.3357902}
\showDOI{\tempurl}


\bibitem[Wu et~al\mbox{.}(2021)]%
        {DynSTGAT}
\bibfield{author}{\bibinfo{person}{Libing Wu}, \bibinfo{person}{Min Wang}, \bibinfo{person}{Dan Wu}, {and} \bibinfo{person}{Jia Wu}.} \bibinfo{year}{2021}\natexlab{}.
\newblock \showarticletitle{DynSTGAT: Dynamic Spatial-Temporal Graph Attention Network for Traffic Signal Control}. In \bibinfo{booktitle}{\emph{Proceedings of the 30th ACM International Conference on Information \& Knowledge Management}} (Virtual Event, Queensland, Australia) \emph{(\bibinfo{series}{CIKM '21})}. \bibinfo{publisher}{Association for Computing Machinery}, \bibinfo{address}{New York, NY, USA}, \bibinfo{pages}{2150–2159}.
\newblock
\showISBNx{9781450384469}
\urldef\tempurl%
\url{https://doi.org/10.1145/3459637.3482254}
\showDOI{\tempurl}


\bibitem[Yu et~al\mbox{.}(2022)]%
        {MAPPO}
\bibfield{author}{\bibinfo{person}{Chao Yu}, \bibinfo{person}{Akash Velu}, \bibinfo{person}{Eugene Vinitsky}, \bibinfo{person}{Jiaxuan Gao}, \bibinfo{person}{Yu Wang}, \bibinfo{person}{Alexandre Bayen}, {and} \bibinfo{person}{Yi Wu}.} \bibinfo{year}{2022}\natexlab{}.
\newblock \showarticletitle{The Surprising Effectiveness of {PPO} in Cooperative Multi-Agent Games}. In \bibinfo{booktitle}{\emph{Thirty-sixth Conference on Neural Information Processing Systems Datasets and Benchmarks Track}}.
\newblock


\bibitem[Zhao et~al\mbox{.}(2012)]%
        {5978226}
\bibfield{author}{\bibinfo{person}{Dongbin Zhao}, \bibinfo{person}{Yujie Dai}, {and} \bibinfo{person}{Zhen Zhang}.} \bibinfo{year}{2012}\natexlab{}.
\newblock \showarticletitle{Computational Intelligence in Urban Traffic Signal Control: A Survey}.
\newblock \bibinfo{journal}{\emph{IEEE Transactions on Systems, Man, and Cybernetics, Part C (Applications and Reviews)}} \bibinfo{volume}{42}, \bibinfo{number}{4} (\bibinfo{year}{2012}), \bibinfo{pages}{485--494}.
\newblock
\urldef\tempurl%
\url{https://doi.org/10.1109/TSMCC.2011.2161577}
\showDOI{\tempurl}


\end{thebibliography}


\clearpage 

\onecolumn
\appendix
\setcounter{figure}{0}
\setcounter{table}{0}
\section{Dynamical traffic flow distributions}\label{appendix:A}
\begin{figure}[h]
  \centering
  \includegraphics[width=0.8\linewidth]{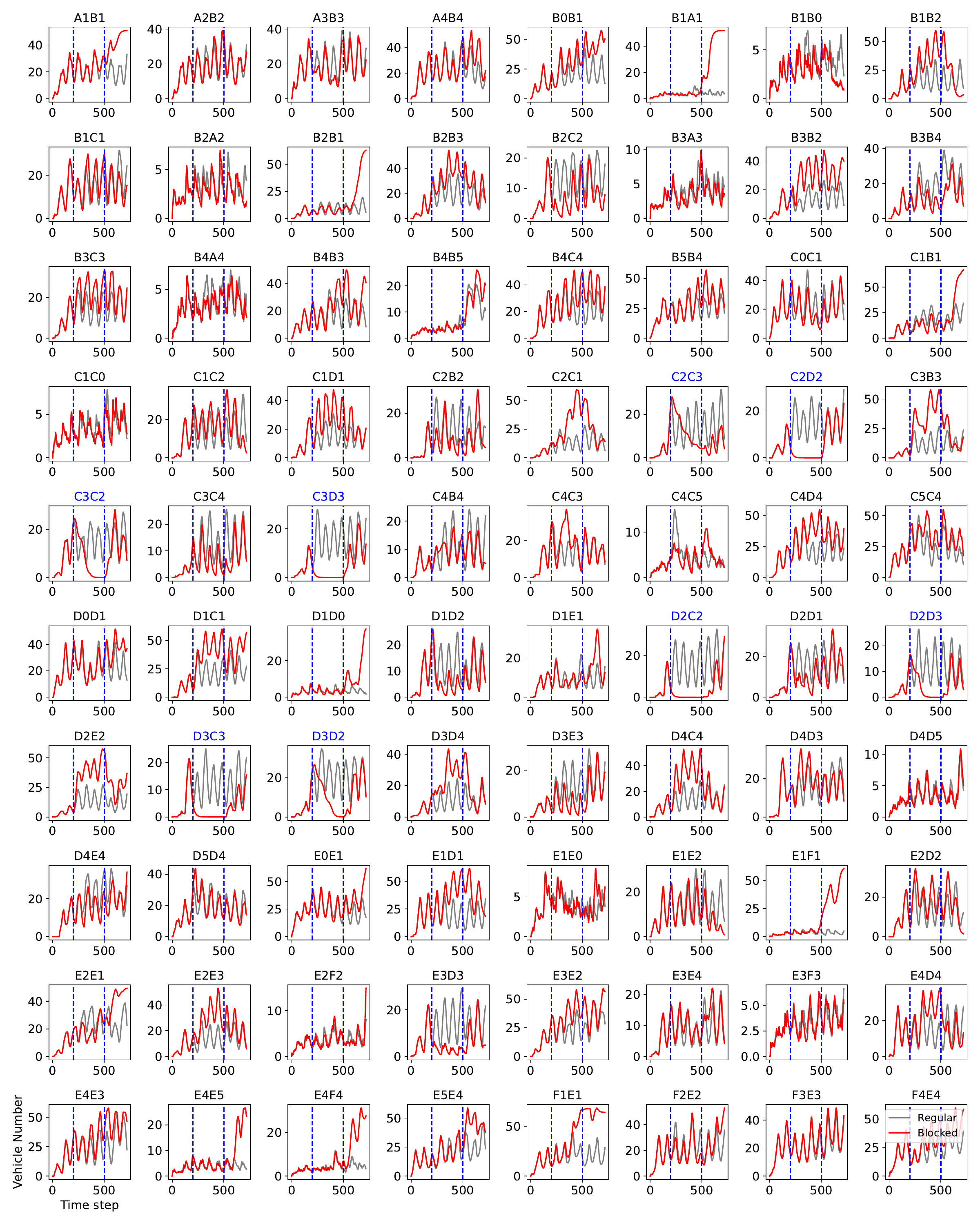}
  \caption{Complete statistics on the changes in the distribution of traffic flow before and after the implementation of road blockage.}
  \label{fig:block_exp}
  \Description{Complete statistics on the changes in the distribution of traffic flow before and after the implementation of road blockage.}
\end{figure}

Referring to Fig. \ref{fig:block_exp}, we conducted two different simulations. The gray curve shows the traffic flow distribution in each lane without any interference to the traffic system, while the red curve shows the traffic flow distribution in each lane with blocking some roads. The blue titles are the numbers of the blocked lanes. The lane blocking time interval is marked with a blue vertical dashed line in the figure. There are cases with opposite signs such as A1B1 and B1A1, indicating two opposite directions of lanes on the same road.

\section{Model parameters}\label{appendix:B}

\begin{table}[htbp]
  \centering
  \caption{MacLight algorithm hyperparameters}
  \begin{tabular}{cccc}
    \toprule
    \textbf{Part} & \textbf{Parameter Name} & \textbf{Value} & \textbf{Description} \\
    \midrule
    \multicolumn{4}{c}{\textbf{PPO}} \\
    \cmidrule(r){1-4}
     & \texttt{actor\_lr}    & 1e-4 & Learning rate for the actor network \\
     & \texttt{critic\_lr}   & 1e-3 & Learning rate for the critic network \\
     & \texttt{lmbda}        & 0.95 & Coefficient for advantage estimation \\
     & \texttt{gamma}        & 0.99 & Discount factor for future rewards \\
     & \texttt{epochs}       & 10   & Number of training epochs \\
     & \texttt{eps}          & 0.2  & Clipping parameter for PPO \\
    \midrule
    \multicolumn{4}{c}{\textbf{PolicyNet and ValueNet Parameters}} \\
    \cmidrule(r){1-4}
    \multirow{4}{*}{\textbf{PolicyNet}} & \texttt{state\_dim} & 33 & Dimension of the input state \\
                                        & \texttt{hidden\_dim} & 66 & Dimension of hidden layers \\
                                        & \texttt{action\_dim} & 8 & Dimension of the output actions \\
                                        & \texttt{layers}      & 3 (fc1, h\_1, fc2) & Number of linear layers \\
    \cmidrule(r){2-4}
    \multirow{5}{*}{\textbf{ValueNet}}  & \texttt{state\_dim}      & 33 & Dimension of the input state \\
                                        & \texttt{hidden\_dim}     & 66 & Dimension of hidden layers \\
                                        & \texttt{global\_emb\_dim} & 16 & Dimension of the global embedding \\
                                        & \texttt{output\_dim}     & 1 & Dimension of the value output \\
                                        & \texttt{layers}          & 3 (fc1, h\_1, fc2) & Number of linear layers \\
    \midrule
    \multicolumn{4}{c}{\textbf{VAE}} \\
    \cmidrule(r){1-4}
     & \texttt{Conv2d layers}        & 3 & Number of convolutional layers in the encoder \\
     & \texttt{ConvTranspose2d layers} & 3 & Number of transposed convolution layers in the decoder \\
     & \texttt{kernel\_size}         & 3 & Size of convolutional kernels \\
     & \texttt{stride}               & 1, 2 & Stride for convolution layers \\
     & \texttt{padding}              & 1 & Padding applied to convolution layers \\
     & \texttt{output\_padding}      & 1 & Output padding in transposed convolution layers \\
     & \texttt{activation}           & ReLU & Activation function used in both encoder and decoder \\
     & \texttt{output\_activation}   & Sigmoid & Output activation function (for normalized image data) \\
     & \texttt{layer\_sizes}         & [64, 128, 256] & Output sizes of each Conv2d layer \\
     & \texttt{convtrans\_layer\_sizes}  & [256, 128, 64] & Output sizes of each ConvTranspose2d layer \\
     & \texttt{flatten\_size}        & 1024 & Size after flattening the encoder output \\ 
     & \texttt{fc\_mu\_size}         & 16 & Output size for the mean in latent space \\
     & \texttt{fc\_logvar\_size}     & 16 & Output size for the log variance in latent space \\
     & \texttt{fc\_decode\_size}     & 1024 & Input size for the decoder from latent space \\ 
    \bottomrule
  \end{tabular}
\end{table}

\newpage
\section{Complete indicators training process}\label{appendix:C}

\begin{figure}[h]
  \centering
  \includegraphics[width=\linewidth]{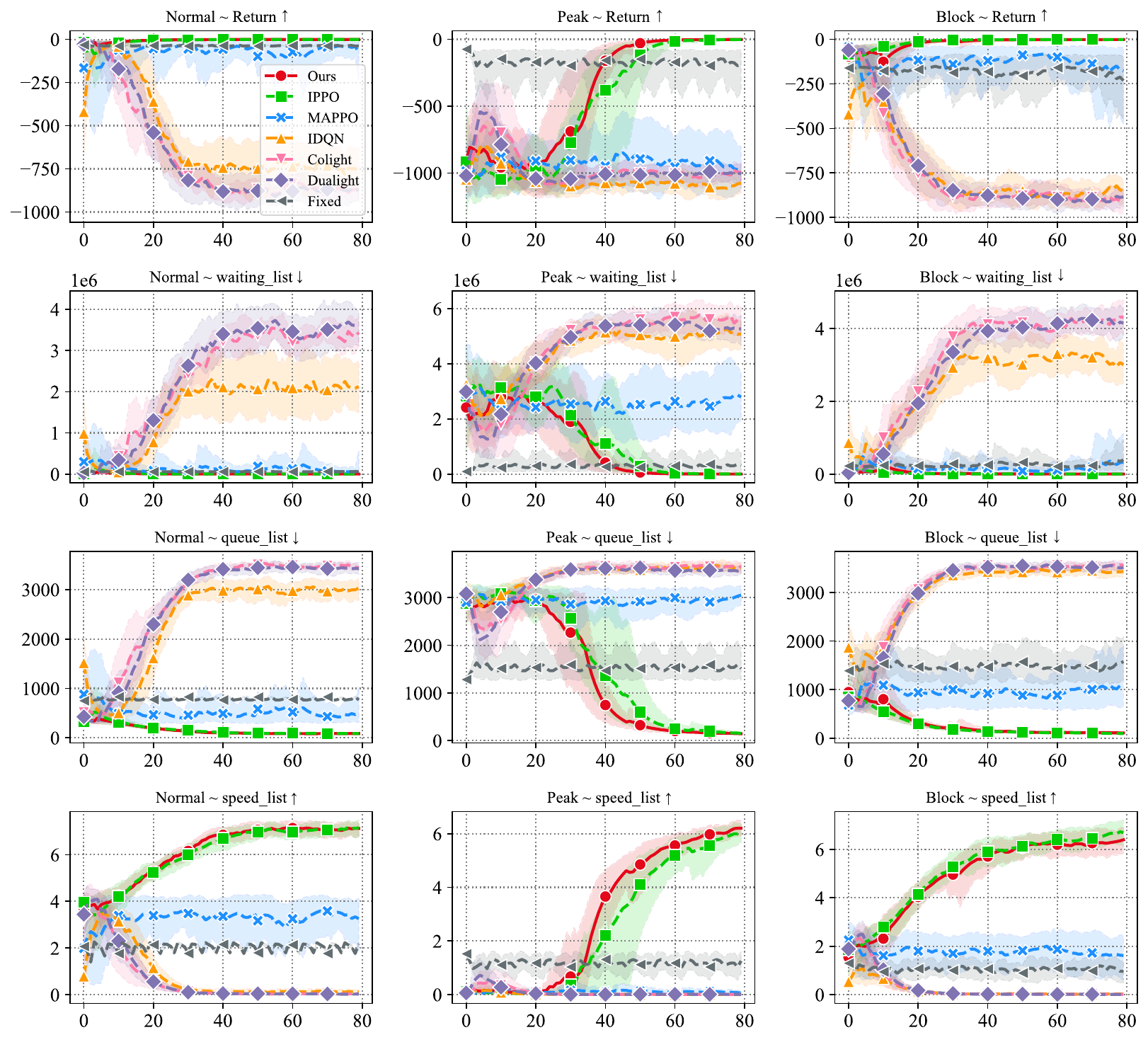}
  \caption{All experimental results. For the Return and Speed indicators, the larger, the better; for the Queue and Waiting indicators, the smaller, the better.}
  \label{fig:total_exp}
  \Description{All experimental results.}
\end{figure}

\end{document}